\documentclass{JamieClass5R}

\usepackage{hyperref}
\usepackage{tagging}
\newcommand{\dd}{g}
\newcommand{\roundUP}{\mathsf{roundUP}}
\newcommand{\roundDOWN}{\mathsf{roundDOWN}}

\usetag{notInWordCount}

\begin{document}
	\title{A Cut-And-Choose Mechanism to Prevent Gerrymandering}
	\author{Jamie Tucker-Foltz\footnote{Amherst College, Amherst, MA, \texttt{jtuckerfoltz@gmail.com}}}
	
	\tagged{notInWordCount}{\maketitle}
	
	\tagged{notInWordCount}{\begin{abstract}
		This paper presents a novel mechanism to endogenously determine the fair division of a state into electoral districts in a two-party setting. No geometric constraints are imposed on voter distributions or district shapes; instead, it is assumed that any partition of the population into districts of equal population is feasible. One party divides the map, then the other party chooses a minimum threshold level of support needed to win a district. Districts in which neither party meets this threshold are awarded randomly. Despite the inherent asymmetry, the equilibria of this mechanism always yield fair outcomes, up to integer rounding.
	\end{abstract}}
	
	\section{Introduction}\label{SectionIntroduction}
	
	Every 10 years, voters in the United States are partitioned into contiguous geographic districts of roughly equal population, according to the national census. During elections, each voter can only vote for a representative in the district of their residence, and the legislature is comprised of one representative per district. By carefully drawing the district boundaries, the party in power can ensure that they receive a disproportionately high fraction of legislative seats. This phenomenon is known as \emph{gerrymandering}, and can have serious political repercussions. For example, in the 2018 midterm elections, in each of Pennsylvania, Michigan, and North Carolina, hundreds of thousands more votes were cast for democratic representatives, yet republicans won more seats in the House of Representatives.\footnote{\raggedright Ingraham, C. 2018. ``In at least three states, Republicans lost the popular vote but won the House." \emph{Washington Post}, November 13. \url{https://www.washingtonpost.com/business/2018/11/13/least-three-states-republicans-lost-popular-vote-won-house}}
	
	At the policy-making level, proposals have been advanced (and some implemented) to prevent gerrymandering, such as requiring that electoral maps satisfy certain shape conditions or a stringent ``symmetry" requirement \cite{Symmetry}, or taking the power out of the hands of politicians completely and  using independent commissions to draw district lines. However, all of these ideas have significant drawbacks. It is easy to find examples of maps that can be drawn with relatively ``simple" shapes, that would appear not to be gerrymandered, but are nevertheless extremely unfair. And with a lack of well-defined rules for fairness, judges have been generally unable to find a clear legal reason to rule any electoral division intentionally biased \cite{LIR}. As for independent commissions, beyond this vague notion of fairness there is not a clear normative consensus as to what the goals of such a commission should be, let alone a system to ensure that they would not act with bias.
	
	This paper adds to the literature on mathematical, rather than legal, solutions to the problem of gerrymandering. It presents a novel mechanism in which two opposing parties endogenously choose an electoral map that satisfies a precise definition of fairness, requiring no further intervention from a neutral third party. It is based on the cut-and-choose paradigm, with two agents, representing the two opposing parties. One agent is arbitrarily assigned to be the ``cutter," and the other to be the ``chooser." The cutter divides the map into electoral districts, then the chooser observes that division and chooses a minimum threshold level of support that the parties need in order to win a given district. For each district in which neither party meets this threshold, the seat is awarded randomly. In all sub-game perfect Nash equilibria of this game, each party ends up winning a number of seats that is proportional to its popular support, up to integer rounding. While there is a possibility for excess randomness, it turns out only to be necessary if the cutter deviates from the equilibrium path.
	
	\subsection{Related literature}\label{SectionLiterature}
	
	Prior mathematical research into gerrymandering falls into two general categories. The first approach aims to detect if a particular map is gerrymandered, so that it can be challenged in court. Tests that rely on the geometry of the districts have been shown to be quite weak. Alexeev and Mixon \cite{GeometryIsInsufficient} show that, under reasonable assumptions about the distribution of voters, the party drawing the districts can turn a 50-50 statewide split into a 70-30 advantage in the legislature, using only ``simple" shapes. Worse yet, if one party has a simple majority, then it is always theoretically possible for them to take \emph{all} of the legislative seats by drawing convex districts \cite{ConvexGerrymandering}. Conversely, legislative maps that are ``fair" are not guaranteed to be simple, and \cite{EfficiencyCriterionCanMakeOddShapes} shows that, in some cases, meeting the ``efficiency criterion" (also called the ``wasted votes criterion," a standard notion of fairness) \emph{requires} complicated district lines, in that the districts may have arbitrarily low Polsby-Popper scores. Given that geometric properties are largely uncorrelated with fairness, methods that do not depend on geometry are more useful. One alternative is to use Markov chains to randomly generate fair district lines. Such methods were used in 2018 in a Pennsylvania Supreme Court case to strike down a congressional map which was determined by a Markov process to be an outlier \cite{Markov}.
	
	The other approach, under which this paper falls, is to rethink the redistricting or election process altogether. One way to guarantee fairness is to have voters vote for parties, rather than representatives. Legislative seats can then be assigned by some apportioning algorithm. Balinski and Young \cite{TheoryOfApportionment} give a formal presentation of this apportionment problem, and \cite{AxiomaticProportionality} axiomatically considers the desirable properties that such an algorithm should have. An election system from the mathematics literature known as \emph{Fair Majority Voting} \cite{FMV} allows voters to vote for specific candidates, while simultaneously keeping track of the votes for each party so that the seats can be apportioned fairly. However, given any electoral map, there is an inevitable high probability that some districts will be won by the candidate with fewer votes, if the apportionment algorithm outputs that a certain party should win more districts than it would have taken otherwise.
	
	In the context of a two-party election, there are also several proposed protocols for drawing electoral districts which do not change the election mechanics, but instead require participation from both parties in the redistricting process. In \cite{K-SplitSolution}, Landau, Reid, and Yershov propose a system in which both parties are presented with a series of different choices over which parts of the map they would like to divide, and the final outcome is determined by finding a point of agreement. However, the menu of choices is determined by a neutral third party, and there are not even theoretical guarantees of how unfair the outcome could be if the neutrality is somehow compromised. The authors acknowledge this flaw, and then attempt to augment their mechanism so that these cases appear less frequently, yet its fairness still hinges on a combination of the third party being non-biased, and random chance. A further drawback is that the maps produced by the mechanism appear ``polarized." Each party ends up with control over specific regions, and they are free to gerrymander those regions to their liking.
	
	Pegden, Procaccia, and Yu \cite{CutAndFreeze} advocate for a protocol called \emph{I-cut-you-freeze}, which elegantly avoids these criticisms. To begin the game, one party draws tentative districts of equal population. Starting with the opposing party, the two parties take turns in which they perform two actions. They first ``freeze" one of the districts which the previous player drew, and then they redraw the remaining unfrozen districts. Play continues until all districts are frozen. Not only does this mechanism hinder the ability of any party to gain a great advantage through gerrymandering, but it also prevents parties from packing any specific population into a single district. One downside of this mechanism is that the districts which get drawn in equilibrium are not completely fair, but still give some extra advantage to the majority party, as well as the party that gets to draw the first map.
	
	Mixon and Villar \cite{UtilityGhost} propose a map drawing game loosely based on the word game Ghost. It allows for exogenously determined units of terrain called \emph{atoms} which cannot be subdivided. Players take turns assigning atoms to districts, and must always ensure that at least one valid complete assignment is still attainable. It is a very flexible mechanism, as the definition of ``valid" can be adjusted as needed. For instance, one can additionally impose the constraint the resulting districts must be connected in the end. In the case where there are no such constraints, the atoms are individual voters, and there are equally as many voters for each party, it is proved that the districts which emerge under optimal play are completely fair, with each party taking half the seats. For more general games with a small number of atoms and different party majorities, the authors demonstrate the mechanism's fairness empirically, by exhaustively solving each game. While there is still a slight unfair bias toward the majority party, they note that the game has less of a first player advantage than other redistricting protocols (such as \cite{CutAndFreeze}).
	
	\subsection{Outline of paper}\label{SectionOutline}
	
	In Section \ref{SectionSetup} the model and proposed mechanism are formally presented, and a real-world example of how the mechanism would work is illustrated through a diagrammatic representation of the game. In Section \ref{SectionResults}, the game is solved and its equilibria characterized. Section \ref{SectionConclusion} concludes by reexamining some of the simplifying assumptions of the model and discussing how the mechanism could be extended to more practical contexts.\\
	
	\section{Redistricting with Threshold Elections}\label{SectionSetup}
	
	\subsection{Setting}\label{SectionSetting}
	
	Consider a setting in which one state, with population $Dn$, must be divided into $D$ districts, each with exactly $n$ voters. There are exactly two political parties, $A$ and $B$, which are the only agents in the game. The utility of each party is the number of districts they win, and both parties are risk-neutral. Every voter has a strict preference over the two parties, and will vote according to their preference. These preferences are common knowledge (this is a standard assumption, and is also used in \cite{K-SplitSolution}, \cite{CutAndFreeze}, and \cite{UtilityGhost}). In total, party $A$ holds a $v_A$ fraction of the total voter population, and party $B$ holds a $v_B = 1 - v_A$ fraction. Call these known, exogenous constants the \emph{party shares} of $A$ and $B$, respectively.
	
	Finally, assume that all possible partitions of voters into $D$ sets of equal size correspond to feasible district divisions. In other words, this model is ``geometry-free," as is common practice in the fair division literature. Some kinds of geometric constraints, and the effect they would have on the mechanism, are briefly considered in Section \ref{SectionConclusion}.
	
	\subsection{Statement of main results}\label{SectionStatement}
	
	The cut-and-choose mechanism proceeds in three steps: a cutting step, a choosing step, and finally, an election.
	\begin{enumerate}
		\item Party $A$ partitions the state into $D$ districts, each with exactly $n$ voters.
		\item Party $B$ observes this division and chooses $m \in [\frac{1}{2}, 1)$.
		\item A special kind of election is held in each district, independently:
		\subitem $\circ$ If either party receives strictly greater than $mn$ votes, that party wins the district.
		\subitem $\circ$ Otherwise, award the district randomly, with equal probability to each party.
	\end{enumerate}
	
	The randomness and asymmetry of this mechanism may seem preposterous at first. It is proved in section \ref{SectionProof} that, on the contrary, there always exists an equilibrium that requires no randomness (beyond tie-breaking), while all equilibria of this game are very fair, up to a small rounding error that becomes insignificant as $D$ rises.
	\begin{theorem}
		If $D > 1$, then there always exists a subgame-perfect Nash equilibrium in which party $B$ sets $m = \frac{1}{2}$.
	\end{theorem}
	\begin{theorem}\raggedright
		In any subgame-perfect Nash equilibrium, in expectation, party $A$ wins a $\roundDOWN(v_A, \frac{1}{2D})$ fraction of the districts, and party $B$ wins a $\roundUP(v_B, \frac{1}{2D})$ fraction of the districts.
	\end{theorem}
	
	Here the functions $\roundUP$ and $\roundDOWN$ round the first argument to an integer multiple of the second argument, i.e.
	\begin{align*}
		\roundUP(a, b) := b\left\lceil \frac{a}{b} \right\rceil, && \roundDOWN(a, b) := b\left\lfloor \frac{a}{b} \right\rfloor.
	\end{align*}
	
	\subsection{An example of how the mechanism works}\label{SectionExample}
	
	The intuitive motivation for this mechanism is as follows. When politicians purposefully gerrymander districts, there are two common tools that they use, called \emph{cracking} and \emph{stacking} (the latter is sometimes also referred to as \emph{packing}). Cracking is when a large population of the opposing party's supporters is spread among multiple districts, so that they barely lose each one, and all votes are wasted. Stacking is when an overwhelming majority of the opposing party's supporters is forced into a single district, wasting that party's excess votes. If party $A$ tries to apply these tactics, it makes them more vulnerable in the choosing stage, as party $B$ can raise $m$ just enough to force the cracked districts party $A$ would have won into the randomized regime, but not enough to affect $B$'s majority in its own stacked districts. Thus, party $A$ has an incentive to divide the districts fairly, so that party $B$ cannot exploit this strategy.
	
	One can see how these dynamics would have played out had the system been used in the 2012 election in Wisconsin, a state well known for partisan gerrymandering. In total, the Democratic candidates (party $B$) won 50.42\% of the vote, but Republicans (party $A$) won 5 of the 8 districts.\footnote{Data taken from \cite{Data}.} In Figure 1, these districts are vertically sorted from most Democratic to most Republican. A zigzagging line is drawn across these district majorities, called the \emph{districting function}, denoted $\dd(m)$. Its vertical segments separate the votes for party $B$ (on the left) from the votes for party $A$ (on the right). For example, the second vertical segment of the districting function indicates that, in the second-most Republican district, the Democrats won just over a 0.37 fraction of the votes.
	
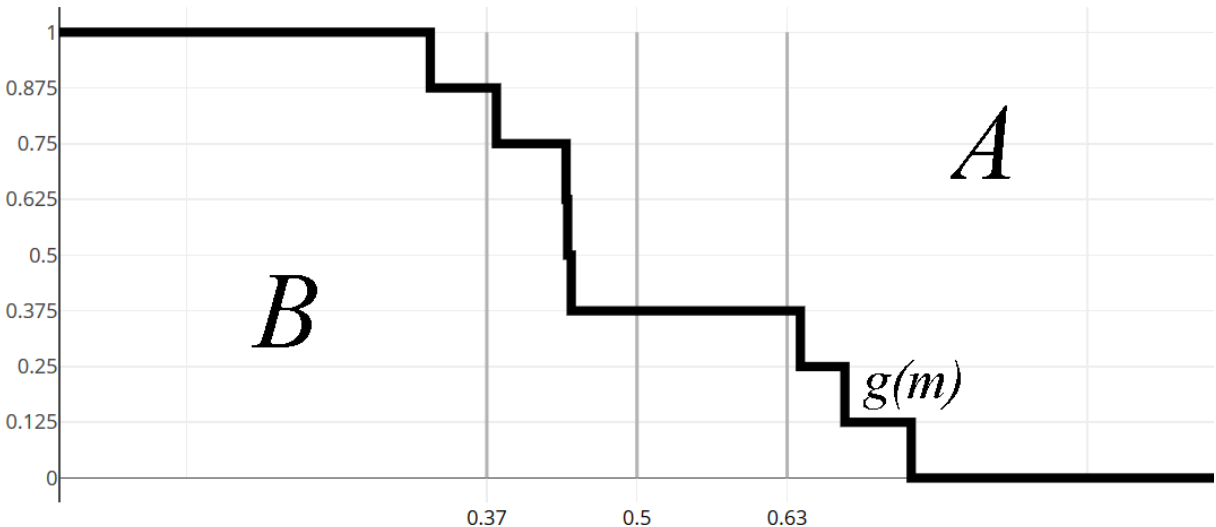
	\tagged{notInWordCount}{\igc{.84}{Graph2-12}{Figure 1. Election results for the 2012 Wisconsin election with $D = 8$ districts. Party $A$ represents the Republicans, and party $B$ represents the Democrats.}}
	
	By representing the data in this form, one can make two important observations. First, given an $m \in [0, 1]$ for which $\dd$ is well defined (i.e. there is not a jump discontinuity), $\dd(m)$ gives the number of districts in which party $B$ received at least an $m$ fraction of votes. Turning the picture upside down, one can analogously see that the number of districts in which party $A$ has a majority greater than $m$ is given by $1 - \dd(1 - m)$. A second observation is that party $B$'s total share can be counted by summing the horizontal strips to the left of the districting function, which is equivalent to summing the area under the curve. If the units are normalized by the total population such that the area of the diagram is 1, then the total fraction of voters supporting party $B$ is
	$$v_B = \int_{0}^{1}\dd(m) dm.$$
	Note that it is possible to compute this integral in more general settings than when $\dd(m)$ is a zigzag line: $\dd(m)$ must always be decreasing and bounded below, so it is always integrable.
	
	This kind of diagram easily explains how these districts appear to be gerrymandered, and how the gerrymandering would have been punished by the threshold voting system. The election system used in the real world is an election where the threshold is $m = 0.5$. That is, party $B$ wins $\dd(0.5) = \frac{3}{8}$ of the districts, and party $A$ wins the other $\frac{5}{8}$ of the districts. Thus, even though party $B$ had a greater party share ($v_B > 0.5$), party $A$ was able to gerrymander the district lines so that it won more districts. However, if party $B$ was able to observe the map and choose a different value of $m$, the outcome will favor party $B$. One such optimal choice is to set $m = 0.63$, as shown on the graph. In that case, $B$ still wins the bottom 3 districts with certainty because $\dd(0.63) = \frac{3}{8}$, but $A$ would win only the top district with certainty because $1 - \dd(1 - 0.63) = \frac{1}{8}$. The remaining 4 districts would be randomly determined, so in expectation, $B$ would win 5 districts, and $A$ would only win 3. As it will be shown in Section \ref{SectionProof}, this is a worse outcome for $A$ than that which would result if $A$ had drawn the district lines more fairly.\\
	
	\section{A Complete Characterization of Equilibria}\label{SectionResults}
	
	\subsection{Proof of main results}\label{SectionProof}
	
	To prove the determinism and fairness results, it is easier to first relax the constraints of party $A$ slightly. Step (1) of the game amounts to choosing the function $\dd(m)$ from a menu of possible zigzagging lines. Suppose instead that it is possible for $\dd(m)$ to be any decreasing function on $[0, 1]$ that starts at $1$ and ends at $0$. In this case it is shown that the equilibria are always completely fair. At the end of the proof, the zigzag constraint for party $A$ is reimposed, which introduces a small rounding error favoring party $B$.
	
	\begin{proof}[Proof of Theorems 1 and 2]
		Let $u_B(\dd, m)$ denote the expected utility of party $B$ (expected number of districts won) given a districting function $g$ and a value for the parameter $m$. There are two pieces that go into the computation of $u_B(\dd, m)$, shown in Figure 2 as the blue and white sections. As discussed in Section \ref{SectionExample}, $B$ will always win a $\dd(m)$ fraction of the districts with certainty (the blue section), and lose a $1 - \dd(1 - m)$ fraction of the districts with certainty (the red section). The remaining $\dd(1 - m) - \dd(m)$ fraction of districts
		will be decided randomly (the white section), so $B$ can expect to win half of those. The expected utility is therefore the sum of the guaranteed wins and the possible wins, with the latter weighted by $\frac{1}{2}$. That is,
		$$\text{ } \ \ \ \ \ \ \ \ \ \ \ \ \ \ \ \ \ \ \ \ u_B(\dd, m) = \dd(m) + \frac{1}{2}[\dd(1 - m) - \dd(m)] = \frac{1}{2}[\dd(m) + \dd(1 - m)]. \ \ \ \ \ \ \ \ \ \ \ \ \ \ \ \ (*)$$
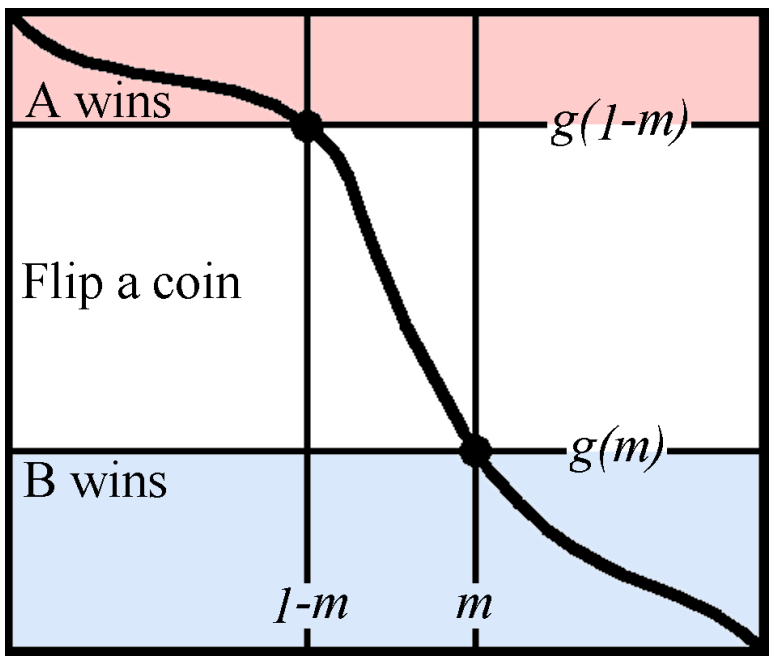
		\tagged{notInWordCount}{\igc{.6}{WhoGetsWhat5}{Figure 2. Visualization of which districts are won by each party, given a districting function $g$ and a specific value of $m$.}}
		We first compute the quantity $\int_{1/2}^{1} u_B(\dd, m) dm$, and show that it does not depend on $\dd$. This fact will allow us to determine an optimal strategy for party $A$. Starting from formula $(*)$,
		\begin{align*}
			\int_{\frac{1}{2}}^{1} u_B(\dd, m) dm &= \int_{\frac{1}{2}}^{1} \frac{1}{2}[\dd(m) + \dd(1 - m)] dm\\
			&= \frac{1}{2}\left(\int_{\frac{1}{2}}^{1} \dd(m)dm + \int_{\frac{1}{2}}^{1} \dd(1 - m) dm\right)\\
			&= \frac{1}{2}\left(\int_{\frac{1}{2}}^{1} \dd(m)dm + \int_{\frac{1}{2}}^{0} \dd(m) (-dm)\right) \stext{sending $m \mapsto 1 - m$}\\
			&= \frac{1}{2}\left(\int_{\frac{1}{2}}^{1} \dd(m)dm + \int_{0}^{\frac{1}{2}} \dd(m) dm\right)\\
			&= \frac{1}{2}\int_{0}^{1} \dd(m) dm\\
			&= \frac{1}{2} v_B.
		\end{align*}
		
		Having established that this integral is constant, we can now solve the game by backward induction. Given a districting function $\dd$ (which party $A$ chose), party $B$ will pick the $m \in [\frac{1}{2}, 1)$ that maximizes $u_B(\dd, m)$. Thus, because the game is zero-sum, party $A$ will choose $g$ to minimize the maximum value of $u_B(\dd, \cdot)$. However, we have seen that no matter what $g$ is, the area under the curve of $u_B(\dd, \cdot)$ over party $B$'s strategy space is fixed at a predetermined value. Therefore, the optimal choice would be to make $u_B(\dd, \cdot)$ be a constant function. Indeed, this is always possible, for one such choice is to set $\dd$ to be constant (there are many other possible choices as well, as will be discussed in Section \ref{SectionEquilibria}).
		
		Given that, in equilibrium, $u_B(\dd, \cdot)$ is constantly equal to some number, say $c$, we know that
		$$\frac{1}{2} v_B = \int_{\frac{1}{2}}^{1} u_B(\dd, m) dm = \int_{\frac{1}{2}}^{1} (c) dm = c\int_{\frac{1}{2}}^{1} (1) dm = \frac{1}{2}c,$$
		so the equilibrium utility is actually $v_B$. Thus, in all equilibria, each party will get utility equal to its party share. Because party $B$ will be indifferent between all values of $m$, setting $m = \frac{1}{2}$ is always one of the optimal choices.
		
		At this point, we have proved Theorems 1 and 2 (without the rounding functions) in the case where $\dd$ is allowed to be an arbitrary decreasing function from $[0, 1]$ to itself. Now suppose $\dd$ is required to actually be a districting function over a finite number of districts $D$. From $(*)$, we see that $u_B$ will necessarily be an integer multiple of $\frac{1}{2D}$, as $\dd$ can only take on values that are multiples of $\frac{1}{D}$. If it so happens that $v_B$ is such a value, then we will still have the same equilibria as in the unrestricted case, as party $A$ can set
		$$\dd(m) = \twocases{m < \frac{1}{2}}{\roundUP(v_B, \frac{1}{D})}{m > \frac{1}{2}}{\roundDOWN(v_B, \frac{1}{D})}.$$
		(Plugging this into $(*)$, one can check that the utility of player $B$ is constant in $m$, always equal to $v_B$, so this is indeed an equilibrium strategy.) If $v_A$ and $v_B$ do not fall into this lattice, then party $A$ will still be able to get the next feasible level of utility, given by $\roundDOWN(v_A, \frac{1}{2D})$. To accomplish this, $A$ can pretend it has less voters than it actually has, to bring $v_B$ up to the nearest point divisible by $\frac{1}{2D}$, then execute any optimal strategy from there. This proves Theorem 2. Finally, to make sure that $m = \frac{1}{2}$ is still an optimal choice for party $B$, party $A$ can just put all the votes they are pretending to give up in either a district in which they have all the votes, or a district in which they are pretending to have none of the votes. This is always possible provided $D > 1$. The effect will be to only make the outcome worse for $B$ if $B$ chooses $m$ sufficiently higher than $\frac{1}{2}$. Therefore, there will still be an equilibrium at $m = \frac{1}{2}$, and Theorem 1 is proved.
	\end{proof}
	
	\subsection{Characterizing equilibria}\label{SectionEquilibria}
	
	What districting functions can arise in equilibrium? Supposing that the party shares fall into the lattice of multiples of $\frac{1}{2D}$, party $A$ must choose $\dd$ such that for all $m \in [\frac{1}{2}, 1)$,
	$$v_B = \frac{1}{2}[\dd(m) + \dd(1 - m)].$$
	By shifting $m$ by $\frac{1}{2}$, multiplying both sides by 2, and rearranging terms, this condition can be rewritten, stating that for all $m \in [0, \frac{1}{2})$,
	$$\left[\dd\left(\frac{1}{2} - m\right) - v_b\right] = -\left[\dd\left(\frac{1}{2} + m\right) - v_b\right].$$
	
	The left hand side gives the vertical difference from $v_b$ to the point on the curve $\dd(m)$ that is $m$ units to the left of center, while the bracketed term on the right hand side gives the difference from $v_b$ to the point on $\dd(m)$ that is $m$ units to the right of center. By enforcing that these two quantities be negatives of each other for all distances $m \in [0, \frac{1}{2})$, it is equivalent to saying that $\dd$ is rotationally symmetric about the point $(\frac{1}{2}, v_b)$. In the case where $v_B = \frac{1}{2}$, this symmetry\footnote{Note that this is an entirely different notion of symmetry than that of the legal proposal described in \cite{Symmetry}. Here, symmetry is with respect to our parameter $m$. King's definition of symmetry is with respect to changing statewide preferences, a dimension which this model does not consider.} criterion precisely captures the meaning of ``fairness" in a district partition: for every district which is stacked or cracked to one party's advantage, by symmetry there is a corresponding district which is stacked or cracked in the equal and opposite way, to the other party's advantage.
	
	When $v_B \neq \frac{1}{2}$, something slightly different happens. As a hypothetical example, suppose there are $D = 4$ districts, where party $A$ has share $v_A = \frac{5}{8}$ and party $B$ has share $v_B = \frac{3}{8}$. These shares are multiples of $\frac{1}{2D} = \frac{1}{8}$, so party $A$ will not have to give up any voters, and under optimal play, both parties will receive expected utilities equal to their party shares. Figure 3 shows three optimal districting functions that $A$ could choose.\\
	
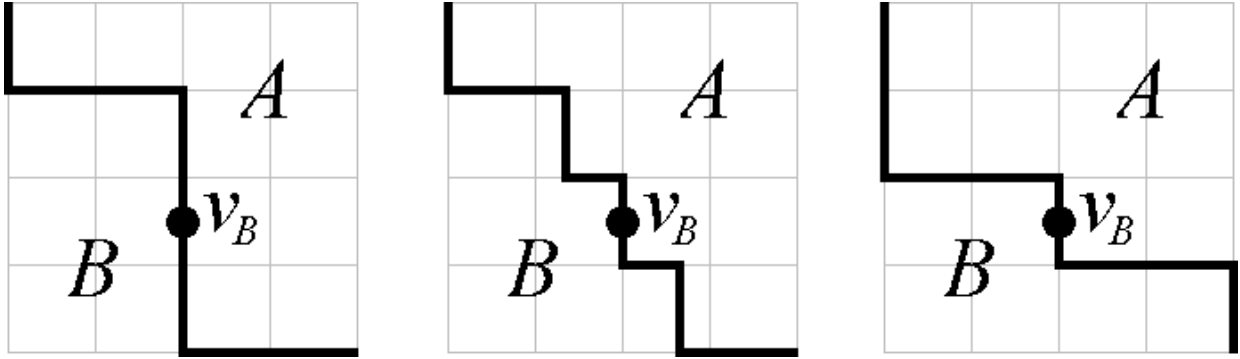
	\tagged{notInWordCount}{\igc{.67}{Graph2-19}{Figure 3. Three optimal districting functions for party $A$, given $D = 4$, $v_A = \frac{5}{8}$, and $v_B = \frac{3}{8}$. There are many other optimal functions.}}
	
	Number the districts 1 to 4 in order from top to bottom. Because $\dd$ must be symmetric about the point $(\frac{1}{2}, \frac{3}{8})$, district 1 must necessarily be completely comprised of voters for party $A$. (Analogously, if party $B$ had a greater share, the point of symmetry would be in the top half of the diagram, and so the bottom district(s) would be completely comprised of voters for party $B$). This makes sense, for otherwise party $B$ could set $m$ higher than $A$'s majority and take at least half of the districts by random chance. Thus, when $v_B \neq \frac{1}{2}$, the rotational symmetry criterion enforces fairness by requiring that there exist districts perfectly stacked with voters for the majority party, reducing any unfair advantage they would otherwise have. This is what makes the mechanism more fair than those presented in \cite{CutAndFreeze} or \cite{UtilityGhost}, in which the majority party inevitably has extra power.
	
	For the bottom three districts, party $A$ has several choices which obey the symmetry criterion. However, to be symmetric about $(\frac{1}{2}, \frac{3}{8})$, there must be a jump in $\dd$ at $\frac{1}{2}$, so district 3 must be comprised of equal amounts of voters for each party.
	
	The scheme on the left divides the remaining space into what might be referred to as ``competitive" districts, in the sense that no party holds a clear majority. Of course, in the present model, voters are perfectly predictable, and the only actual factor of randomness is that there will be a tie in each of these three districts, so the winner will be decided by coin flip. Notice that $B$ will win 1.5 districts in expectation, which is a $\frac{3}{8}$ fraction of districts, consistent with Theorem 2.
	
	On the right is a maximally ``safe" division, as districts 2 and 4 will each be won unanimously. Regardless of the choice of $m$, party $B$ will win district 4 with certainty, and district 3 with probability $\frac{1}{2}$, again for an expected value of 1.5 districts.
	
	Between these two extremes lies another arbitrary possibility, the middle diagram, with a districting function that has intermediate jump points across districts 2 and 4. This effectively gives $B$ a choice over the variance. If they set $m$ below the final jump point they will get the safe outcome, and if they set $m$ above the final jump point they will get the random outcome. However, because the graph is symmetric about $(\frac{1}{2}, \frac{3}{8})$, districts 2 and 4 will be thrown into the random regimes at the exact same value of $m$, so $B$ cannot gain any real advantage. The expected utility in all cases will be $\frac{3}{8}$.\\
	
	\section{Extensions}\label{SectionConclusion}
	
	While this mechanism fairly allocates districts in a stylized two party setting, there are numerous extensions and practical considerations that merit further attention. The equilibria described seem to be contingent on somewhat unreasonable assumptions about the real world, such as universal domain over districting functions or perfect information. Even under these assumptions, the mechanism has several shortcomings, most notably an unavoidable ``rounding" unfairness to party $A$, and a mere existence guarantee of maximally-deterministic equilibria. Below are some of the more important and mathematically intriguing directions for future research.
	
	\textbf{Imperfect stacking.}
	The reliance of the mechanism on perfect information and unrestricted districting may seem like a serious flaw, as any equilibrium where $\roundDOWN(v_A, \frac{1}{2D}) \neq \roundUP($ $v_B, \frac{1}{2D})$ requires that party $A$ perfectly stack one of the majority party's districts, so that there are no voters supporting the minority party. If $A$ has a greater share of the population, and even one voter for party $B$ sneaks into each of its stacked districts, then $B$ can set $m \geq \frac{n - 1}{n}$ and expect to take half of the seats. Thus, if the domain of districting functions precludes perfect stacking, or there is no way to predict votes with certainty, the deterministic equilibrium which gives party $A$ a fair, comfortable advantage breaks down into complete randomness, with party $A$ losing its edge.
	
	With a slight tweak, the mechanism can avoid such disastrous consequences. Suppose that there is some fundamental limit $M \in [\frac{1}{2}, 1)$ above which no district can be stacked to a greater majority. In this case party $B$ should be required to choose $m \in [\frac{1}{2}, M)$, rather than any $m$. That way, party $A$ does not have to fear $B$ taking advantage of its inability to perfectly stack, and the outcome will be as fair as possible.
	
	For this simple model of the kinds of limitations that politicians face when drawing the map, there is a simple modification to the mechanism that preserves a level of fairness. There are several factors one could add to make the model richer. However, no matter what model one constructs, the way to adapt the mechanism should still follow the same general idea: whatever natural restrictions party $A$ faces in district drawing should be met with prescribed restrictions on the choice $B$ gets to make in step 2.
	
	\textbf{Reducing randomness.}
	One simple modification to reduce randomness would be to equally distribute the districts in which neither party reached a majority of more than $m$. Only if there is an odd number of such districts would it be necessary to randomize the very last district. The equilibria of the game would be unchanged, as both parties are still getting the same expected utility from these districts. While this solution eliminates, or at least significantly reduces, the variance in state-wide utilities for each party, there might still be randomness within individual districts that both voters and politicians would be irked by. A better kind of solution would be one in which $m = \frac{1}{2}$ is a \emph{unique} equilibrium. For example, one could specify that while $B$ needs a majority of $m$ to take a district with certainty, $A$ needs only a majority of $\frac{m}{2} + \frac{1}{4}$, so $B$ would be obliged to keep $m$ low. However, if one tries to incentivize this choice in step 2, party $A$ may end up exploiting this incentive in step 1, bringing $B$ back to the point of indifference, and the net effect will have been to give an arbitrary advantage to party $A$, while $m = \frac{1}{2}$ may still not be a unique equilibrium. Such incentives must be applied more subtly.
	
	\textbf{Counteracting the rounding unfairness.} There may be settings in which the rounding-error advantage of party $B$ simply disappears. For instance, any form of imperfect information has the effect of smoothing out the vertical pieces of the districting function. It is quite possible that, in the presence of a certain kind of noise, it will always be possible for party $A$ to choose a districting function that is (mostly) symmetric about $(\frac{1}{2}, v_B)$. If the rounding unfairness is not naturally eliminated, there may still be ways to reward or punish parties arbitrarily, to balance the utilities of the cutter and the chooser.
	
	\textbf{Abstaining voters.} A final obstacle is the existence of abstaining voters, who count in the census for the purposes of equal populations in each district, but do not contribute to either party's voter share. One possible modification would be to introduce another parameter to the election mechanics, similar to $m$, which prescribes a required amount of voter participation needed for either party to be able to win outright. However, it is not clear exactly how such a parameter would need to be chosen. Understanding the impact of non-voters on this kind of mechanism is not just important in its own right, but might also be a necessary step toward extending the mechanism to a setting with more than two parties.\\
	
	\tagged{notInWordCount}{\section*{Acknowledgments}
	
	Special thanks to Brian Baisa for his very generous advice and feedback. Also thanks to Christopher Kingston, Ariel Procaccia, and several members in my audience at the The 29th International Conference on Game Theory at Stony Brook University for helpful questions and comments.\\}
	
	\tagged{notInWordCount}{\begin{flushleft}\end{flushleft}}

\begin{thebibliography}{99}
		\bibitem{GeometryIsInsufficient}
		Alexeev, Boris, and Dustin G. Mixon. 2017. ``Partisan gerrymandering with geographically compact	districts." arXiv:1712.05390. \url{https://arxiv.org/pdf/1712.05390}
		
		\bibitem{EfficiencyCriterionCanMakeOddShapes}
		Alexeev, Boris, and Dustin G. Mixon. 2018. ``An Impossibility Theorem for Gerrymandering." \textit{The American Mathematical Monthly} 125 (10): 878–84. doi:10.1080/00029890.2018.1517571.
		
		\bibitem{FMV}
		Balinski, Michel L. 2008. ``Fair Majority Voting (or How to Eliminate Gerrymandering)." \textit{The American Mathematical Monthly} 115 (2): 97–113. doi:10.1080/00029890.2008.11920503.
		
		\bibitem{AxiomaticProportionality}
		Balinski, Michel L., and G. Demange. 1989. ``An Axiomatic Approach to Proportionality Between Matrices." \textit{Mathematics of Operations Research} 14 (4): 700–719. doi:10.1287/moor.14.4.700.
		
		\bibitem{TheoryOfApportionment}
		Balinski, Michel L., and H. Peyton Young. 1982. \textit{Fair Representation: Meeting the Ideal of One Man, One Vote}. New Haven, CT: Yale University Press.
		
		\bibitem{Markov}
		Chikina, Maria, Alan Frieze, and Wesley Pegden. 2017. ``Assessing Significance in a Markov Chain without Mixing." \textit{Proceedings of the National Academy of Sciences} 114 (11): 2860–64. doi:10.1073/pnas.1617540114.
		
		\bibitem{LIR}
		Gottlieb, Stephen. 2017. ``Gerrymandering in the Supreme Court." Lecture presented at Amherst, MA., October 22. 
		
		\bibitem{Symmetry}
		Grofman, Bernard, and Gary King. 2007. ``The Future of Partisan Symmetry as a Judicial Test for Partisan Gerrymandering after LULAC v. Perry." \textit{Election Law Journal: Rules, Politics, and Policy} 6 (1): 2–35. doi:10.1089/elj.2006.6002.
		
		\bibitem{K-SplitSolution}
		Landau, Z., O. Reid, and I. Yershov. 2008. ``A Fair Division Solution to the Problem of Redistricting." \textit{Social Choice and Welfare} 32 (3): 479–92. doi:10.1007/s00355-008-0336-6. 
		
		\bibitem{UtilityGhost}
		Mixon, Dustin. G., and Soledad Villar. 2018. ``Utility Ghost: Gamified redistricting with partisan symmetry." arXiv:1812.07377. \url{https://arxiv.org/pdf/1812.07377}
		
		\bibitem{CutAndFreeze}
		Pegden, Wesley, Ariel Procaccia, and, Dingli Yu. 2017. ``A partisan districting protocol with provably nonpartisan outcomes." arXiv:1710.08781. \url{https://arxiv.org/pdf/1710.08781}
		
		\bibitem{ConvexGerrymandering}
		Sober\'on, Pablo. 2017. ``Gerrymandering, Sandwiches, and Topology." \textit{Notices of the American Mathematical Society} 64 (09): 1010–13. doi:10.1090/noti1582.
		
		\bibitem{Data}
		Vaughn, Christy, Sachet Bangia, Bridget Dou, Sophie Guo,; and Jonathan Mattingly. 2016. ``Quantifying Gerrymandering." Information Initiative at Duke. \url{https://services.math.duke.edu/~sb337/gerrymandering/WI.html}
	\end{thebibliography}
\end{document}